\documentclass[12pt,preprint]{aastex}

\slugcomment{
KUNS~1820
}

\newcommand{\beq}{\begin{equation}}
\newcommand{\beqa}{\begin{eqnarray}}
\newcommand{\eeq}{\end{equation}}
\newcommand{\eeqa}{\end{eqnarray}}

\newcommand{\bm}[1]{\mbox{{\boldmath $#1$}}} 

\begin{document}

\title{  
Cosmological X-ray Flashes in the Off-Axis Jet Model
}

\author{Ryo Yamazaki\altaffilmark{1}}
\affil{Department of Physics, Kyoto University, Kyoto 606-8502, Japan}
\email{yamazaki@tap.scphys.kyoto-u.ac.jp}

\and

\author{Kunihito Ioka\altaffilmark{2}}
\affil{Department of Earth and Space Science, Osaka University, 
Toyonaka 560-0043, Japan}
\email{ioka@vega.ess.sci.osaka-u.ac.jp}

\and

\author{Takashi Nakamura\altaffilmark{3}}
\affil{Department of Physics, Kyoto University, Kyoto 606-8502, Japan}
\email{takashi@yukawa.kyoto-u.ac.jp}

\def\d{{\rm d}}
\def\p{\partial}
\def\w{\wedge}
\def\o{\otimes}
\def\f{\frac}
\def\tr{{\rm tr}}
\def\Half{\frac{1}{2}}
\def\half{{\scriptstyle \frac{1}{2}}}
\def\T{\tilde}
\def\RA{\rightarrow}
\def\N{\nonumber}
\def\n{\nabla}
\def\bb{\bibitem}
\def\BE{\begin{equation}}
\def\EE{\end{equation}}
\def\BEA{\begin{eqnarray}}
\def\EEA{\end{eqnarray}}
\def\L{\label}
\def\VVM{\langle V/V_{\rm max}\rangle}
\def\zero{{\scriptscriptstyle 0}}
\begin{abstract}
The $\VVM$ of the cosmological 
X-ray flashes  detected by Wide Field Cameras on {\it BeppoSAX}
is calculated theoretically in  a simple jet model.
The total emission energy from the jet is assumed to be constant.
We find that if the jet opening half-angle is smaller 
than 0.03 radian, 
the theoretical  $\VVM$ for fixed opening half-angle 
is less than $\sim0.4$,
which is consistent with the recently reported observational value of
$0.27\pm 0.16$ at the 1 $\sigma$ level.
This suggests  that the off-axis GRB jet with the  small
opening half-angle at the cosmological distance can be identified
as the cosmological X-ray flash.
\end{abstract}


\section{INTRODUCTION}
The X-ray flash (XRF) is a class of X-ray transients
(Heise et al. 2001, see also Barraud et al. 2003).
Some properties of XRFs, such as the observed event rate, 
the duration and the isotropic distribution, are similar to that of 
Gamma-Ray Bursts (GRBs),
while the spectral hardness of XRFs characterized by 
the peak flux ratio, the fluence ratio and the photon index is
softer than that of GRBs.
This class represents a large portion of the whole GRB population.
Recently, the observational value of $\VVM$,
which is the measure of the homogeneity of spatial distribution
(Schmidt, Higdon, \& Hueter 1988; 
see also Chang \& Yi 2001; Kim, Chang, \& Yi 2001),
has been updated
from $0.56\pm0.12$ (J.~Heise 2000, talk given in 
2nd workshop Gamma-Ray Bursts in the Afterglow Era)
to $0.27\pm0.16$ (J.~Heise 2002, talk given in 
3rd workshop Gamma-Ray Bursts in the Afterglow Era).
The updated value of $\VVM$ suggests that XRFs take place at
a cosmological distance.

Various models accounting for the nature of the XRFs
have been proposed
(Yamazaki, Ioka, \& Nakamura 2002; Heise et al. 2001; 
Dermer, Chiang, \& B\"{o}ttcher 1999; Huang, Dai, \& Lu 2002;
M\'{e}sz\'{a}ros et al. 2002;
Mochkovitch et al. 2003; Daigne \& Mochkovitch 2003).
Heise et al. (2001) proposed that XRFs could be 
{\it GRBs at high redshift}.
The redshifts of  XRF011030 and XRF020427 have 
an upper limit of $z\lesssim 3.5$
(Bloom et al. 2003; 
J.~Heise 2002, talk given in 
3rd workshop Gamma-Ray Bursts in the Afterglow Era), 
which suggests that XRFs take place at not so high redshift but
the same as that of GRBs.
{\it The photosphere-dominated fireball model}
may account the nature of the XRFs with peak energy $E_p$
more than $\sim20$ keV (M\'{e}sz\'{a}ros et al. 2002,
Ramirez-Ruiz \& Lloyd-Ronning 2002).
However, further considerations are needed to explain the event
with $E_p\sim $ of approximately a few keV, 
such as 
XRF020427 (L.~Amati 2002, talk given in 
3rd workshop Gamma-Ray Bursts in the Afterglow Era),
XRF020903 and XRF010213 (N.~Kawai 2002, talk given in 
3rd workshop Gamma-Ray Bursts in the Afterglow Era).
The models with small Lorentz factors, such as
{\it the dirty fireball model} (Dermer, Chang, \& B\"{o}ttcher 1999; 
Huang, Dai, \& Lu 2002) or 
{\it the structured-jet model} (Rossi, Lazzati, \& Rees 2001;
Woosley et al. 2002; Zhang \& M\'{e}sz\'{a}ros 2002a)
also have possibilities to explain the properties of the XRF,
with the implicit assumption that the XRF arises 
not from internal shocks (Zhang \& M\'{e}sz\'{a}ros 2002b). 
The models for internal shocks with 
{\it small contrast of high Lorents factors}
might be the origin of XRFs
(Mochkovitch et al. 2003; Daigne \& Mochkovitch 2003).

For the other possibility,
we have studied {\it the off-axis jet model} and proposed
that if we observe the GRB jet with a large viewing angle,
it looks like an XRF (Yamazaki, Ioka, \& Nakamura 2002,
Yamazaki, Ioka, \& Nakamura 2003).
In Yamazaki, Ioka, \& Nakamura (2002)  the value of 
the jet opening half-angle was  adopted as  $\Delta\theta=0.1$.
In this  model  the distance to the
farthest XRF ever detected is about 2 Gpc ($z\sim0.4$) 
 so that the cosmological effect is small and 
$\VVM\sim 0.5$.
Recent observations suggest that GRBs with relatively 
small opening angle exist, while 
the distribution of $\Delta\theta$ is not yet clear
(Panaitescu \& Kumar 2002).
If we assume the total emission energy to be constant
as in the previous paper,
the intrinsic luminosity is larger for the smaller opening half-angle.
Such GRBs at the cosmological distance 
observed from off-axis viewing angle 
may be seen as XRFs 
and $\VVM$ is expected to be  smaller than 0.5.

In this paper, we will show that our off-axis model has a
possibility of accounting for the observational value of $\VVM$
if we change some of the model parameters in the previous paper
 (Yamazaki, Ioka, \& Nakamura 2002).
This paper is organized as follows.
In \S~\ref{sec:model}, we describe a simple jet model 
including the effect of cosmological expansion.
We assume the uniform jet with sharp edges.
Although one may consider the structured jet 
motivated by the simulation of the collapsar model,
we cannot conclude, both observationally and theoretically,
which model is preferable.
The $\VVM$ for the XRFs detected by the Wide Field Cameras (WFCs) 
on {\it BeppoSAX} is calculated in \S~\ref{sec:vvm}.
Section~\ref{sec:dis} is devoted to a discussion.
Throughout this paper, 
we adopt the flat universe with 
$\Omega_M=0.3$, $\Omega_\Lambda=0.7$ and $h=0.65$.

\section{EMISSION MODEL OF X-RAY FLASHES}\label{sec:model}
We consider a simple jet model of XRFs in the previous
papers (Yamazaki, Ioka, \& Nakamura 2002; Ioka \& Nakamura 2001)
taking into account  the cosmological effect.
A general formula to calculate the observed flux  
from an optically thin material is derived by 
Granot, Piran, \& Sari (1999) and Woods \& Loeb (1999).
Here we adopt their formulations and notations.
Let us use a spherical coordinate system ${\bm r}=(r, \theta, \phi)$
in the central engine frame, where the $\theta=0$ axis points 
toward the detector and the central engine is located at the origin.
Consider a photon emitted at time $t$ and place ${\bm r}$ 
in the central engine frame. 
It will reach the detector at a time $T$ given by
\begin{equation}
T = (1+z)T_{z} = (1+z)(t -{r \mu/ c}) \ ,
\label{time}
\end{equation}
where $\mu \equiv \cos \theta$ and 
$z$ is the cosmological redshift of the source;
$T=0$ was chosen as the time of arrival at the detector of a
photon emitted at the origin at $t=0$.
Then the observed flux at the observed time $T$ and 
observed frequency $\nu$, measured in 
erg s$^{-1}$ cm$^{-2}$ Hz$^{-1}$, is given by
\begin{equation}
F_\nu(T)=\f{1+z}{d_L^2}\int^{2\pi}_0d\phi \int^1_{-1}d\mu
\int^{\infty}_0r^2dr\,
\f{j'_{\nu'} \left(\Omega_d', {\bm r}, T_z+{r\mu/c}\right)}
{\gamma^2(1- \beta \mu)^2}\ ,
\label{flux;j}
\end{equation}
where $d_L$, $\Omega_d'$ and $j'_{\nu'}$ 
are the luminosity distance to the source,
the direction towards the detector measured in the frame
comoving with the jet (comoving frame),
and the comoving frame emissivity 
in units of ergs s$^{-1}$ cm$^{-3}$ Hz$^{-1}$ sr$^{-1}$.
The frequency $\nu'$, which is measured in the comoving frame,
is given by 
\begin{equation}
\nu'=\nu_z\gamma(1-\beta\mu) = (1+z)\nu\gamma(1-\beta\mu)\ .
\label{frequency}
\end{equation}

We adopt an instantaneous emission of infinitesimally thin shell 
at $t=t_0$ and $r=r_0$.
If the emission is isotropic in the comoving frame,
the emissivity has a functional form of
\begin{eqnarray}
j'_{\nu'}(\Omega_d', {\bm r}, t)&=&A_0 f(\nu') \delta(t-t_0) 
\delta(r-r_0) H(\Delta \theta-|\theta-\theta_v|)\N\\
&&\times H\left[\cos \phi-\left({{\cos\Delta \theta
-\cos\theta_v \cos\theta}\over
{\sin \theta_v \sin\theta}}\right)\right],
\end{eqnarray}
where  $f(\nu')$ and $A_0$ represent the spectral shape and the
amplitude, respectively.
$\Delta\theta$ and $\theta_v$ are the jet opening half-angle 
and the viewing angle 
that the axis of the emission cone makes
with the $\theta=0$ axis. 
The delta functions describe
an instantaneous emission at $t=t_0$ and $r=r_0$,
and $H(x)$ is the Heaviside step function which
describes that the emission is inside the cone.
Then the observed flux of a single pulse is given by
\begin{eqnarray}
F_{\nu}(T)
=\f{2(1+z)r_0 c \gamma^2 A_0}{d_L^2}
{{\Delta \phi(T) f\left[\nu_z\gamma (1-\beta\cos\theta(T))\right]
}\over{\left[\gamma^2 (1-\beta\cos\theta(T))\right]^2}},
\label{eq:jetthin}
\end{eqnarray}
where, $1-\beta\cos\theta(T)=(1+z)^{-1}({c\beta}/{r_0})(T-T_0)$
and $T_0=(1+z)(t_0-r_0/c\beta)$.
For $\Delta \theta>\theta_v ~{\rm and}~ 0<\theta(T)\le \Delta
\theta - \theta_v$, $\Delta \phi(T)=\pi$,
otherwise 
$\Delta \phi(T)=
\cos^{-1}\left[
{{\cos \Delta \theta - \cos \theta(T) \cos \theta_v}\over
{\sin \theta_v \sin \theta(T)}}\right]$.
For $\theta_v < \Delta \theta $, $\theta(T)$ varies from 0 to 
$\theta_v+\Delta \theta$ while from $\theta_v-\Delta \theta$ to
$\theta_v+\Delta\theta$ for $\theta_v > \Delta \theta $. 
In the latter case, $\Delta \phi(T)=0$ for $\theta(T)=\theta_v-\Delta
\theta$.
Pulse-starting  and ending time are given by
\begin{eqnarray}
T_{start}&=&T_0+(1+z)({r_0}/{c\beta})
(1-\beta\cos(\max[0,\theta_v-\Delta\theta])),\N\\
T_{end}&=&T_0+(1+z)({r_0}/{c\beta})
(1-\beta\cos(\theta_v+\Delta\theta)).
\end{eqnarray}

The observed spectrum of GRBs is well approximated by 
the Band spectrum (Band et al. 1993). In order to 
have a spectral shape similar to the Band spectrum,
we adopt the following form of the spectrum in the comoving frame,
\begin{equation}
f(\nu')=\left\{
\begin{array}
{c@{, \ \ }c@{}}
(\nu'/\nu'_0)^{1+\alpha_B} & \nu'<\nu'_0 \\
(\nu'/\nu'_0)^{1+\beta_B} & \nu'>\nu'_0
\end{array}
\right.
\label{eq:spectrum}
\end{equation}
where $\alpha_B$ ($\beta_B$) is the low (high) energy power law index.
Equations (\ref{eq:jetthin}) and (\ref{eq:spectrum})
are the basic equations to calculate the flux of a single pulse.

In order to study the dependence on the viewing angle 
$\theta_v$ and the jet opening half-angle $\Delta\theta$,
we fix the other parameters as
$\alpha_B=-1$, $\beta_B=-3$, 
$\gamma\nu'_0=200\,{\rm keV}$, $r_0/c\beta \gamma^2=10\,{\rm s}$,
and $\gamma=100$ (Preece et al. 2000).
We fix the amplitude $A_0$ so that
the isotropic $\gamma$-ray energy 
$E_{iso}=4\pi d_L^2 (1+z)^{-1}S(20-2000\,{\rm keV})$ satisfies
\begin{equation}
\f{(\Delta\theta)^2}{2}E_{iso}
=5\times10^{50}{\rm ergs},
\end{equation}
when $\theta_v=0$ and $z=1$ (Frail et al. 2001).
Here $S(\nu_1-\nu_2)=
\int_{T_{start}}^{T_{end}}F(T;\nu_1-\nu_2)dT$
is the fluence in the energy range $\nu_1-\nu_2$
and $F(T;\nu_1-\nu_2)=\int_{\nu_1}^{\nu_2}
F_\nu(T)d\nu$ is the flux in the same energy range.
The values of $A_0$ for different opening angles are
summarized in Table~\ref{TableVVM}.
When the jet opening half-angle $\Delta\theta$ becomes smaller,
$A_0$ becomes larger. 

\section{CALCULATION OF $\VVM$}\label{sec:vvm}
\begin{figure}
\plotone{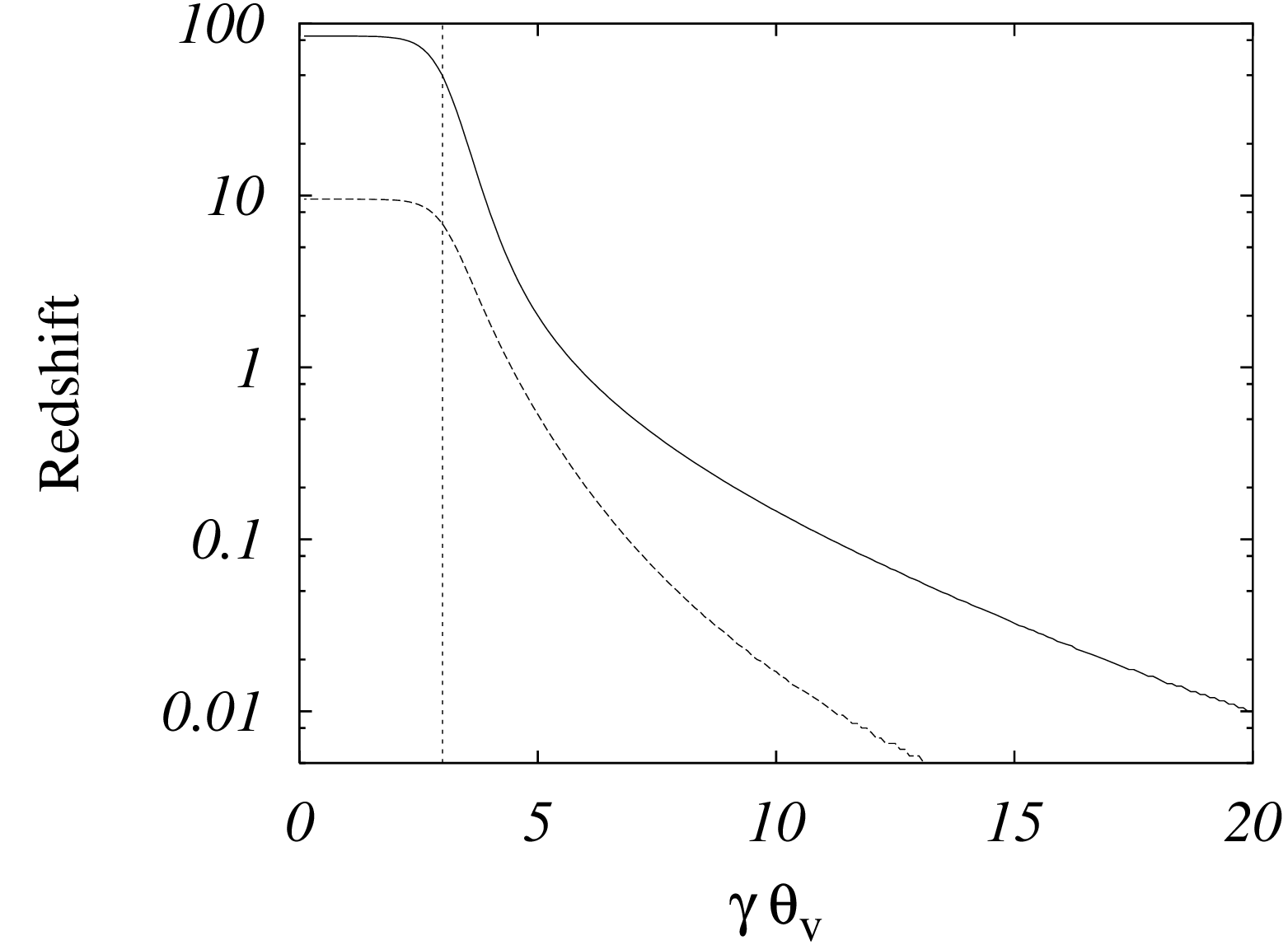}
\caption{
Maximum and  minimum redshift of the XRF as a function
of the viewing angle $\gamma\theta_v$ are shown
in the case of $\Delta\theta=0.03$.
The solid line (dashed line) represents the maximum (minimum) redshift
$z_{max}$ ($z_{min}$).
The jet emission is observed as the XRF if the source has a redshift
$z$ in the range $z_{min}<z<z_{max}$ (see text).
The vertical dashed line represents $\theta_v=\Delta\theta=0.03$.
}\label{fig_redshift}
\end{figure}

The formalism to calculate $\VVM$ is given by
Mao \& Paczy\'{n}ski (1992) and Piran (1992).
In our case the absolute luminosity and spectrum depend on
 $\theta_v$ and $\Delta\theta$.
The observed-integrated number count from the
source at redshift $z$ with the jet opening half-angle 
$\Delta\theta$ and the viewing angle $\theta_v$ is given by
\begin{equation}
C(\Delta\theta,\theta_v,z)=\int_{T_{start}}^{T_{end}}dT
\int_{\nu_1}^{\nu_2}
d\nu \f{F_\nu(T,\Delta\theta,\theta_v,z)}{h\nu}.
\label{count}
\end{equation}
Let $z_{min}$  and $z_{max}$ be the minimum and the maximum redshift 
of the  XRF for given $\Delta\theta$ and $\theta_v$.
In determining  $z_{min}$  and $z_{max}$, we should  note 
that the operational definition of the XRF detected by 
{\it BeppoSAX} is the fast X-ray
transient with duration less than $\sim10^3$ seconds which is
detected by WFCs and not detected by the 
Gamma-Ray Burst Monitor (GRBM).
Therefore, if the sources are nearby such that $z<z_{min}$,
they are observed as GRBs because the observed fluence
in the $\gamma$-ray band ($40-700$ keV) becomes larger than
the limiting sensitivity of GRBM 
($\sim 3\times10^{-6}{\rm ergs}\ {\rm cm}^{-2}$).
If the sources are too far such that $z>z_{max}$,
they cannot be observed by WFCs, which have an observation band
of $2-28$ keV and a limiting sensitivity
of about $4\times10^{-7}{\rm ergs}\ {\rm cm}^{-2}$.
Although the detection conditions of instruments vary with
many factors of each event, such as the duration, the spectral index,
or the peak photon energy (Band 2002), 
we adopt very simple criteria here.
As shown in Figure~\ref{fig_redshift}, 
both $z_{max}$ and $z_{min}$ depend on $\theta_v$ and $\Delta\theta$.

For given $z_{max}$, we can calculate
the minimum integrated count 
$C_{min}=C(\Delta \theta, \theta_v, z_{max})$.
Then, $\VVM$ for given $\Delta\theta$ and $\theta_v$ 
is calculated as
\begin{equation}
\left<\f{V}{V_{max}}\right>_{\Delta\theta,\,\theta_v}=
\f{\int_{z_{min}}^{z_{max}}[C(z)/C_{min}]^{-3/2}\,{\cal D}(z)\,dz}
{\int_{z_{min}}^{z_{max}}{\cal D}(z)\,dz},
\label{vvm1}
\end{equation}
where ${\cal D}(z)$ is given as
\begin{equation}
{\cal D}(z)=\f{n(z)}{1+z}\,4\pi \left(\f{d_L}{1+z}\right)^2\,
\f{d}{dz}\left(\f{d_L}{1+z}\right).
\label{eq:Dz}
\end{equation}
The luminosity distance is given by
\begin{equation}
d_L(z)=\f{c}{H_0}(1+z)\int_0^z\f{dz'}
{\sqrt{\Omega_\Lambda+\Omega_M(1+z')^3}}.
\end{equation}
The function $n(z)$ is the comoving GRB rate density.
We assume that $n(z)$ is proportional to the comoving rate densities
of the star formation $n_{SF}(z)$ in the following form
(Madau \& Pozzetti 2000; Porciani \& Madau 2001)
\begin{equation}
n_{SF}(z)=0.46h\f{\exp(3.4z)}{\exp(3.8z)+45}
\sqrt{\Omega_M + \Omega_\Lambda(1+z)^{-3}}
\ M_\sun\ {\rm yr}^{-1}\ {\rm Mpc}^{-3} .
\end{equation}

When we consider the case of $z_{min}<z_{max}\ll 1$,
the cosmological effect can be neglected, so that
${\cal D}\propto z^2$ and
$C/C_{min}\sim (z/z_{max})^{-2}$.
Then, Eq.~(\ref{vvm1}) becomes
\begin{equation}
\left<\f{V}{V_{max}}\right>_{\Delta\theta,\,\theta_v} =
0.5\left[1+(z_{min}/z_{max})^3 \right] .
\end{equation}
Therefore, $z_{min}\ll z_{max}\ll 1$ implies
$\VVM_{\Delta\theta,\,\theta_v} \sim 0.5$.

Next, we integrate Eq.~(\ref{vvm1}) over $\theta_v$ as
\begin{equation}
\left<\f{V}{V_{max}}\right>_{\Delta\theta}=
\f{\int
\VVM_{\Delta\theta,\,\theta_v}
\, W(\theta_v)\,d\theta_v}
{\int W(\theta_v)\,d\theta_v},
\label{vvm3}
\end{equation}
where $W(\theta_v)$ is the weight function
which is the product of the solid angle factor and
the volume factor:
\begin{equation}
W(\theta_v)=2\pi\sin\theta_v
\int_{z_{min}}^{z_{max}}{\cal D}(z)\,dz .
\end{equation}
The results of the numerical integration
are summarized in Table 1.
For each $\Delta\theta$,
$z_{max}(\theta_{v,\, p})$ ($z_{min}(\theta_{v,\, p})$)
means the maximum (minimum) redshift at which $W(\theta_v)$ takes the
maximum value.
If we take the jet opening half-angle as 
$\Delta\theta\lesssim0.03$, $\VVM_{\Delta\theta}$ is smaller 
than $\sim0.4$, which is consistent with the observational 
result at the  1 $\sigma$ level.

\begin{figure}
\plotone{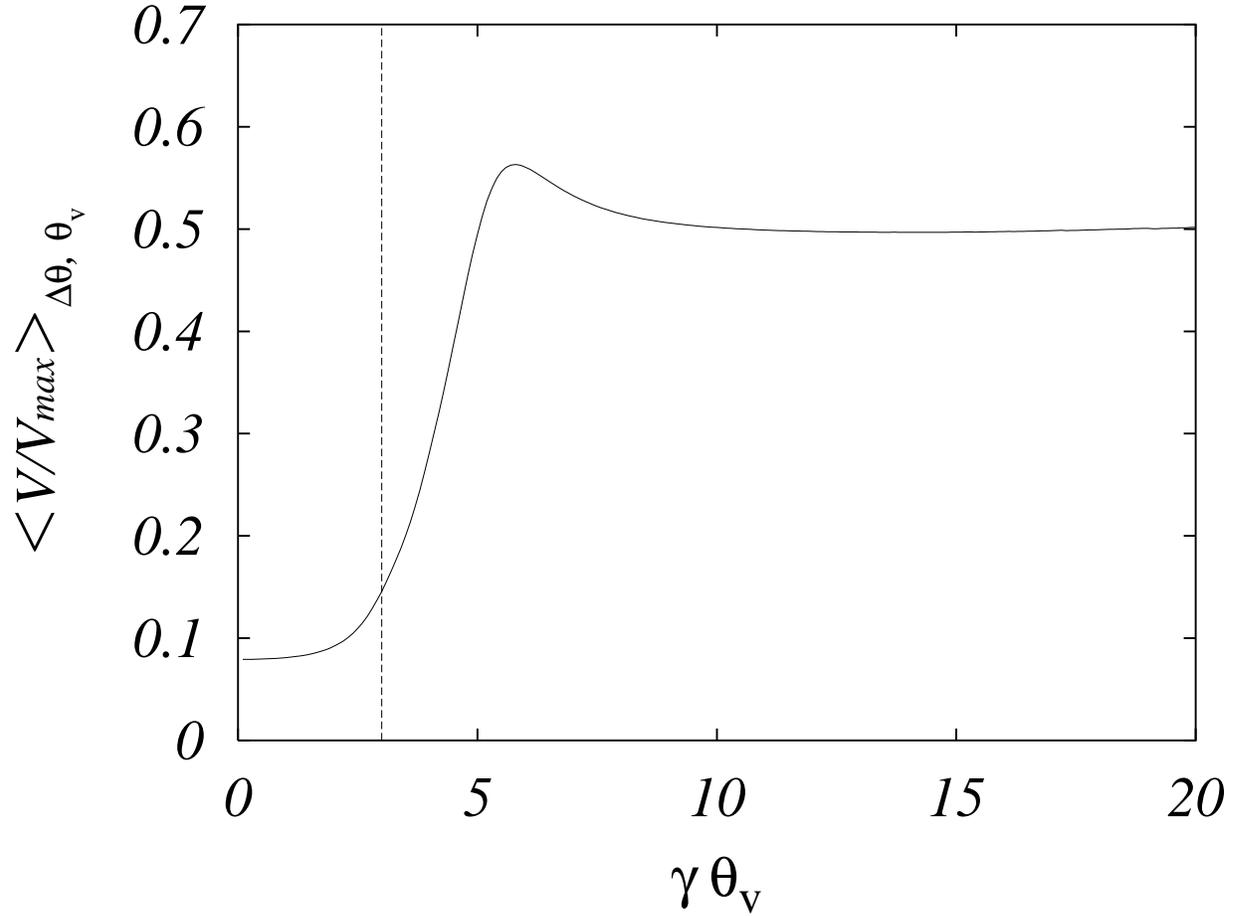}
\caption{
$\VVM_{\Delta\theta,\,\theta_v}$ for the XRF detected by the 
WFCs on {\it BeppoSAX} is shown as a function of
the viewing angle $\gamma\theta_v$ in the case of $\Delta\theta=0.03$.
}\label{fig_vvm}
\end{figure}

Let us consider the behavior of $\VVM_{\Delta\theta,\,\theta_v}$
in the case of the fixed opening half-angle $\Delta\theta=0.03$.
Recent analysis of the GRB afterglows shows that some GRBs
have a jet with a opening half-angle of less than 0.05 radian, i.e.,
the smallest value of $\Delta\theta$ is about 0.03
(Panaitescu \& Kumar, 2002).
Let us decrease the viewing angle $\theta_v$ 
from a sufficiently large value ($\gamma\theta_v\sim20$).
As shown in Figure~\ref{fig_redshift}, 
both $z_{max}$ and $z_{min}$ increase monotonically, 
since the observed flux from the source increases
due to the relativistic beaming effect.
However, the behavior of $\VVM_{\Delta\theta,\,\theta_v}$
is more complicated since the function
${\cal D}(z)$ in equation (\ref{eq:Dz})
has the maximum value at $z=z_p\sim1.5$.
We plot $\VVM_{\Delta\theta,\,\theta_v}$ as a function of
$\gamma\theta_v$ in Figure~\ref{fig_vvm}.
If $\theta_v$ is large enough   $z_{max}$ is smaller
than $z_p$. 
Then the cosmological effect is small,
so that $\VVM_{\Delta\theta,\,\theta_v}\sim 0.5$.
While if $\theta_v$ is small enough  $z_{min}$
is larger than $z_p$. Then ${\cal D}(z)$ is a decreasing function 
in the range $z_{min}<z<z_{max}$,
and the contribution of XRFs at smaller distance  to
$\VVM_{\Delta\theta,\,\theta_v}$ is larger 
so that $\VVM_{\Delta\theta,\,\theta_v}$ is small.
We note  that the above  argument  does not depend on
 $\Delta\theta$ so much.

\begin{figure}
\plotone{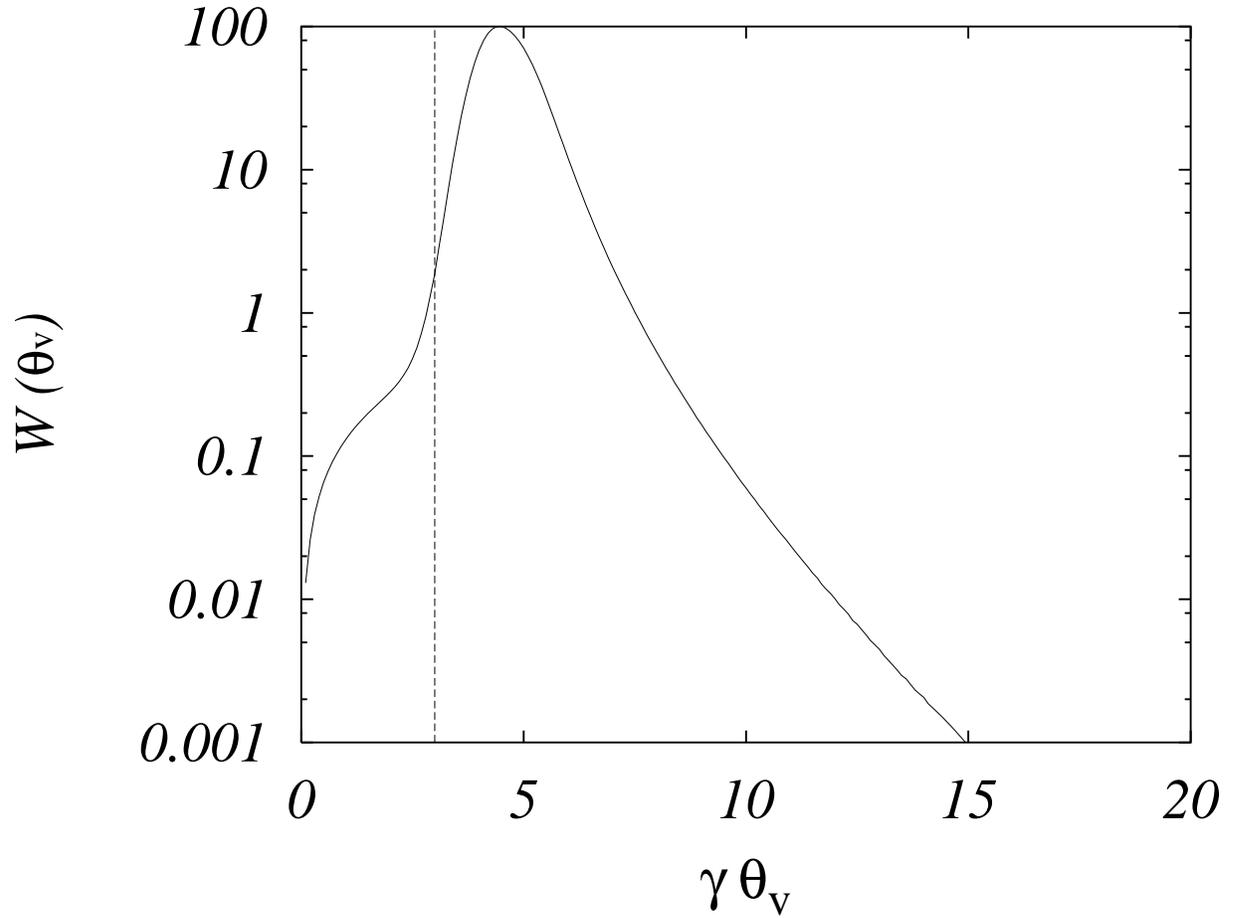}
\caption{
The weight function $W(\theta_v)$, 
which is the relative observed event rate,
is shown as a function of the viewing angle $\gamma\theta_v$.
Note that the normalization of $W(\theta_v)$ is arbitrary.
The vertical dashed line represents $\theta_v=\Delta\theta=0.03$.
}\label{fig_weight}
\end{figure}

The behavior of the weight function $W(\theta_v)$
is shown in Figure~\ref{fig_weight}.
When $\theta_v$ is large enough for $z_{max}$ to be smaller
than $z_p$ or when $\theta_v$ is small enough,
$W(\theta_v)$ is small since ${\cal D}(z)$ 
or the solid angle factor are
relatively small in the range $z_{min}<z<z_{max}$.
One can see that in the case of $\Delta\theta=0.03$,
$W(\theta_v)$ takes the maximum value at 
$\gamma\theta_{v,\, p}\sim4.5$, when $z_{min}\sim z_p$.

Above discussions show that the sources 
with the viewing angle $\theta_v\sim0.05$ at $z\sim1.5$
are the most frequent class of the XRFs in the population for
the opening half-angle $\Delta\theta=0.03$.
In other cases in which $\Delta\theta$ is smaller than 0.03,
the typical values are $z\sim1.5$ and 
$\theta_v\sim \Delta\theta+0.02$ since
$W(\theta_v)$ takes maximum at $\theta_{v,\,p}\sim\Delta\theta+0.02$
(see Table~\ref{TableVVM}).

\section{DISCUSSION}\label{sec:dis}
We have calculated $\VVM_{\Delta\theta}$ 
for the emission from a simple jet model,
and shown that when the jet opening half-angle $\Delta\theta$
is smaller than about 0.03, $\VVM_{\Delta\theta}$ 
for the XRFs detected by WFCs on {\it BeppoSAX} is smaller than 0.4.
The value of $\Delta\theta\sim0.03$ 
has been obtained from the fitting of the afterglow light curve
(Panaitescu \& Kumar 2002; Frail et al. 2001).
Such a narrow jet which is inferred 
by afterglow observations is rare.
However, following Eq.~(1) of Frail et al. (2001),
the jet break time is given by 
$t_j\sim 13(\Delta\theta/0.01)^{8/3}$~minutes
and so it requires fast localization to observe the jet break
for a narrow jet.
Therefore, at present, the small number of GRBs with small
$\Delta\theta$ may come from the observational selection effect.
In the context of this scenario, we might be able to account 
for the fact that
afterglows of XRFs have been rarely observed since the afterglow 
at a fixed time gets dimmer for an earlier break time.
Furthermore, some ``dark GRBs''
might be such a small opening angle jet observed
with an on-axis viewing angle for the same reason.

We briefly comment on how the results obtained in this paper
will depend on the Lorentz factor of the jet $\gamma$.
We see that when $\gamma$ becomes large, $\VVM_{\Delta\theta}$
becomes small.
For example, when we fix $\gamma=200$, we obtain
$\VVM_{\Delta\theta=0.03}=0.39$ and
$\VVM_{\Delta\theta=0.1}=0.43$.
This implies that the limitation on $\Delta\theta$ can be relaxed.

We can estimate the typical observed photon energy as 
$h\nu_{obs}\sim (1+z)^{-1}\delta\nu'_0,$
where $\delta^{-1}=\gamma[1-\beta\cos\tilde{\theta}]$
and $\tilde{\theta}={\rm max}\{0,\,\theta_v-\Delta\theta\}$
(Yamazaki, Ioka, \& Nakamura 2002).
Since $\gamma\gg 1$ and $\theta_v,\,\Delta\theta\ll 1$,
we obtain 
\begin{equation}
h\nu_{obs}\sim \f{2\gamma\nu'_0}{(1+z)
[1+(\gamma\tilde{\theta})^2]}.
\end{equation}
In \S~\ref{sec:vvm}, 
we have shown that for fixed $\Delta\theta\lesssim0.03$,
the typical value of $\theta_v$ is $\sim\Delta\theta+0.02$.
Therefore, for the adopted parameters $\gamma\nu'_0=200$keV and
the typical redshift $z=z_p\sim 1.5$, 
one can derive $h\nu_{obs}\sim 30$ keV,
which is the typical observed peak energy of the XRFs
(Kippen et al. 2002).
We can propose from our argument that
the emissions from the jets with a small opening half-angle 
such as $\Delta\theta\lesssim0.03$ are 
observed as XRFs when they are seen from off-axis viewing angle.
If one can detect the afterglow of the XRF,
which has the maximum flux at about several hours after the XRF,
the fitting of light curve may give us the key information
about the jet opening angle (Granot et al. 2002).
Therefore, our theoretical model can be tested by the near-future
observations. 

We can estimate the observed event rate of the XRF 
for fixed $\Delta\theta$ as,
\begin{equation}
R_{{\rm XRF},\,\Delta\theta}=\f{1}{4\pi}\int W(\theta_v)\,d\theta_v.
\end{equation}
In order to calculate $R_{{\rm XRF},\,\Delta\theta}$,
we consider the proportionality constant 
${\cal R}=n(z)/n_{SF}(z)$.
One can write approximately as ${\cal R}=r_{>8M_\sun}k$,
where $r_{>8M_\sun}$ is the number of stars with masses
$M>8M_\sun$ per unit mass.
Since we assume all stars with masses $M>8M_\sun$
explode as core-collapse SNe,
$k$ represents the ratio of the number of XRF sources
to that of core-collapse SNe.
We adopt the value $k=1\times10^{-3}$,
which is derived by the result of Porciani \& Madau (2001)
combined with the effect of the solid angle factor
$(\Delta\theta)^2/2$.
Using a Salpeter initial mass function $\phi(M)$, we obtain
$r_{>8M_\sun}=(\int_{8M_\sun}^{125M_\sun}\phi(M)\,dM)/
(\int_0^{125M_\sun}M\phi(M)\,dM)
=1.2\times10^{-2}M_\sun^{-1}$.
Then, in the case of $\Delta\theta=0.03$, we derive
$R_{{\rm XRF},\,\Delta\theta}=1\times 
10^2 \ {\rm events} \ {\rm yr}^{-1}
({\cal R}/1\times10^{-5}M_\sun^{-1})$,
which is comparable to the observed event rate of the XRF
(Heise et al. 2001).
Note that the value of $R_{{\rm XRF},\,\Delta\theta}$ remains 
unchanged within 
a factor of 2 when we vary $\Delta\theta$ from 0.01 to 0.07.

When the jet opening half-angle $\Delta\theta$ has a distribution
$f_{\Delta \theta}$,
we integrate $\VVM_{\Delta\theta}$ and $R_{{\rm XRF},\,\Delta\theta}$ 
over the distribution of $\Delta\theta$ as
\begin{equation}
\VVM=
\f{\int d(\Delta\theta)\,
f_{\Delta\theta}\,R_{{\rm XRF},\,\Delta\theta}\,
\VVM_{\Delta\theta}}
{\int d(\Delta\theta)\,f_{\Delta\theta}\,
R_{{\rm XRF},\,\Delta\theta}}
\ \ ,
\end{equation}
\begin{equation}
R_{{\rm XRF}}=\f{\int d(\Delta\theta)\,f_{\Delta\theta}\,
R_{{\rm XRF},\,\Delta\theta}}
{\int d(\Delta\theta)\,f_{\Delta\theta}}
\ \ ,
\end{equation}
respectively.
We assume a power-low distribution as
$f_{\Delta\theta}\propto(\Delta\theta)^{-q}$.
When we adopt $q=4.54$ (Frail et al. 2001), and 
integrate over $\Delta\theta$ from 0.01 to 0.2 rad,
we find $\VVM=0.36$ and 
$R_{{\rm XRF}}=1\times10^2$~events~yr$^{-1}$.
These values mainly depend on the lower bound of the integration.
For example, we obtain $\VVM=0.43$ and 
$R_{{\rm XRF}}=3$~events~yr$^{-1}$ if the integration is done 
over $\Delta\theta$ from 0.03 to 0.2 rad
(Note that we may let the value of $R_{\rm XRF}$ be
consistent with observed value by adjusting ${\cal R}$).
Since the statistics of the observational data will increase
in the near future owing to instruments such as
{\it HETE-2} and {\it Swift},
we will be able to say more than above discussion, 
including more accurate functional form of $f_{\Delta\theta}$
than that we have considered above,
as well as the relation to the GRB event rate.


\acknowledgments
We would like to thank the referee for useful comments and
suggestions.
We are grateful to L.~Amati for valuable comments.
Numerical computation in this work was carried out at the 
Yukawa Institute Computer Facility.
This work was supported in part by
Grant-in-Aid for Scientific Research 
of the Japanese Ministry of Education, Culture, Sports, Science
and Technology, No.05008 (R.Y.), No.00660 (K.I.), 
No.14047212 (T.N.), and No.14204024 (T.N.).

\appendix



\clearpage

\begin{deluxetable}{cccccc}
\tablecaption{ RESULTS OF CALCULATION FOR FIXED $\Delta\theta$
\label{TableVVM}
}
\tablewidth{0pt}
\tablecolumns{2}
\tablehead{
\colhead{$\Delta\theta$} & 
\colhead{$A_0$\tablenotemark{a}} &
\colhead{$\theta_{v,\, p}$\tablenotemark{b}} & 
\colhead{$z_{max}(\theta_{v,\, p})$} & 
\colhead{$z_{min}(\theta_{v,\, p})$} & 
\colhead{$\VVM_{\Delta\theta}$\tablenotemark{c}} 
}
\startdata
0.10 & 0.84 & 0.103 & 2.8 & 1.5 & 0.46 \\
0.09 & 1.0 & 0.095 & 2.9 & 1.4 & 0.45 \\
0.08 & 1.3 & 0.086 & 3.0 & 1.4 & 0.44 \\
0.07 & 1.7 & 0.077 & 3.1 & 1.3 & 0.44 \\
0.06 & 2.3 & 0.068 & 3.3 & 1.2 & 0.44 \\
0.05 & 3.4 & 0.060 & 3.5 & 1.2 & 0.44 \\
0.04 & 5.2 & 0.052 & 3.6 & 1.1 & 0.43 \\
0.03 & 9.3 & 0.045 & 3.8 & 0.99  & 0.40 \\
0.02 & 22 & 0.038 & 4.0 & 0.89 & 0.38 \\
0.01 & 109 & 0.034 & 4.1 & 0.77 & 0.35 \\
\enddata
\tablenotetext{a}{In units of 
${\rm erg}\ {\rm cm}^{-2}\ {\rm Hz}^{-1}$.}
\tablenotetext{b}{The viewing angle at which the weight function
$W(\theta_v)$ takes the maximum value.}
\tablenotetext{c}{For the XRFs detected by WFCs on BeppoSAX.}
\end{deluxetable}

\end{document}